\documentclass[a4paper, amsfonts, amssymb, amsmath, reprint, showkeys, nofootinbib, twoside]{revtex4-1}
\usepackage[english]{babel}
\usepackage[utf8]{inputenc}
\usepackage{color}
\def\1#1{{\color{black}#1}}
\usepackage{amsthm}
\usepackage{mathtools}
\usepackage{xcolor}
\usepackage{graphicx}
\usepackage[left=23mm,right=13mm,top=35mm,columnsep=15pt]{geometry} 
\usepackage{adjustbox}
\usepackage{placeins}
\usepackage[T1]{fontenc}
\usepackage{csquotes}
\usepackage[pdftex, pdftitle={Article}, pdfauthor={Author}]{hyperref} 
\begin{document}
\title{Linear chirp instability analysis for ultrafast pulse metrology}

\author{Esmerando Escoto}
    \thanks{These authors contributed equally to this work.}
    \affiliation{Max Born Institute for Nonlinear Optics and Short Pulse Spectroscopy, Max-Born-Stra{\ss}e 2A, 12489 Berlin, Germany. \\
    Correspondence email: escoto@mbi-berlin.de}

\author{Rana Jafari}
    \thanks{These authors contributed equally to this work.}
    \affiliation{School of Physics, Georgia Institute of Technology, 837 State Street, Atlanta, GA, USA }
    
\author{G\"unter Steinmeyer}
    \affiliation{Max Born Institute for Nonlinear Optics and Short Pulse Spectroscopy, Max-Born-Stra{\ss}e 2A, 12489 Berlin, Germany}
    \affiliation{Physics Department, Humboldt University, Newtonstra{\ss}e 15, 12489 Berlin, Germany}
\author{Rick Trebino}
    \affiliation{School of Physics, Georgia Institute of Technology, 837 State Street, Atlanta, GA, USA}

\date{\today} 

\begin{abstract}
Pulse train instabilities have often given rise to confusion in misinterpretation in ultrafast pulse characterization measurements. Most prominently known as the coherent artifact, a partially mode-locked laser with non-periodic waveform may still produce an autocorrelation that has often been misinterpreted as indication for a coherent pulse train. Some modern pulse characterization methods easily miss the presence of a coherent artifact, too. Here we address the particularly difficult situation of a pulse train with chirp-only instability. This instability is shown to be virtually invisible to autocorrelation measurements, but can be detected with FROG, SPIDER, and dispersion scan. Our findings clearly show that great care is necessary to rule out a chirp instability in lasers with unclear mode-locking mechanism and in compression experiments in the single-cycle regime. Among all dynamical pulse train instabilities analyzed so far, this instability appears to be the best hidden incoherence and is most difficult to detect.
\end{abstract}


\maketitle

\section{Introduction}

In the past two decades, numerous techniques have been developed for the full characterization of the intensity and phase of ultrashort pulses  \cite{trebino2012frequency,iaconis1998spectral,Lozovoy2004,Stibenz2005,miranda2012simultaneous, wollenhaupt2012short}. Compared to the more traditional second-order autocorrelation measurement \cite{Diels1985}, full characterization methods not only deliver precise pulse durations, but they can also resolve the pulse shape, that is, structure in the temporal or spectral intensity and phase of potentially complex pulses \cite{Xu2008,escoto2018advanced}. However, most such techniques operate multi-shot, so they inherently assume stability of the pulse shape in the pulse train.  Worse, there is no ``pulse-shape stability meter,'' so the task of determining the pulse-shape stability also necessarily falls to the pulse-measurement technique. If the pulse intensity or phase varies on a time scale shorter than the measurement time, a misleading narrow temporal structure arises in such a measurement that is commonly referred to as the ``coherent artifact,'' first pointed out by Fisher and Fleck for intensity autocorrelation \cite{fisher1969phase} and recently studied in detail for modern methods by Rhodes {\em et al.} \cite{rhodes2013pulse,rhodes2014standards,rhodes2017unstable}. \1{This coherent artifact is effectively the autocorrelation of the transform limit}
\begin{equation}
\chi(t) = \int_0^\infty  \left<\tilde{E}_j(\omega) \tilde{E}_k^*(\omega) \right>_{j=k} \exp(i \omega t) {\rm d}\omega .
\end{equation}
Here $\tilde{E}_j(\omega)$ is the \1{oscillating} electric field \1{(i.e., without slowly-varying envelope approximation)} of the $j$-th pulse in the train, and $<\dots >$ refers to ensemble averaging. \1{$\chi(t)$ relates to the pulse-to-pulse coherence, which} can be described by \cite{dudley2006supercontinuum}
\begin{equation}
\Gamma(\omega) =  \frac{\left<\tilde{E}_j(\omega) \tilde{E}_k^*(\omega) \right>_{j \ne k}}{\left<\tilde{E}_j(\omega) \tilde{E}_k^*(\omega) \right>_{j = k}}.
\end{equation}
In the situation of a strongly degraded pulse-to-pulse coherence ($\Gamma(\omega)$ much smaller than unity), intensity autocorrelations tend to measure the coherent artifact. \1{In the worst-case scenario of a multimode continuous-wave laser, one measures the autocorrelation of $\chi(t)$, i.e., the coherent artifact, even though there is no mode-locking mechanism present \cite{kornaszewski2012sesam, Wilcox2013, kornaszewski2013reply}. In the situation of partial mode-locking, one observes a coherence spike on top of the autocorrelation of an embedding pulse structure} \cite{ratner2012coherent}. In particular, a degraded interpulse coherence appears difficult or even impossible to detect.  Additional measurements of the radio frequency spectrum of the pulse train are common, but this is more appropriate for simple pulse-energy fluctuations and does not indicate the detailed pulse-shape fluctuations that can occur. Otherwise, continuously operating multimode lasers can be erroneously interpreted as being mode-locked \cite{Wilcox2013}. Indeed, the situation of irregular bursts of short pulses can be difficult to detect in autocorrelations when the durations of the coherent artifact and the actual pulse autocorrelation function are within an order of magnitude \cite{rhodes2013pulse, ratner2012coherent}.

For the situation of irregular bursts of short pulses, the coherent-artifact problem can be detected by some, but not all full characterization methods. Frequency-resolved optical gating (FROG) has proven adept at revealing the different types of instabilities studied so far \cite{rhodes2013pulse,rhodes2014standards,rhodes2017unstable}. Discrepancies between the measured and retrieved FROG traces reliably indicate an unstable pulse train \cite{rhodes2013pulse, ratner2012coherent}. This is due to the unique relationship between a field and its FROG trace (up to trivial ambiguities) \cite{trebino2012frequency,bendory2017uniqueness}. Additionally, it has been demonstrated that FROG can measure partially coherent pulses by only modifying the processing of its measurement. For instance, the XUV pulses from free-electron lasers that are randomly delayed with respect to the expected synchronized infrared pulses (within a given temporal jitter envelope) can be retrieved from the jitter-averaged traces \cite{bourassin2015partially} and such an approach could be generalized to other sources of decoherence \cite{bourassin2015partially,escoto2019retrieving}. Multiphoton intrapulse interference phase scan (MIIPS) was shown to be able to quantify this instability \cite{rhodes2017unstable,lozovoy2015quantifying}. However, other methods have been reported to miss the presence of a coherent artifact, namely spectral phase interferometry for direct electric-field reconstruction (SPIDER) \cite{rhodes2013pulse,ratner2012coherent}, and self-referenced spectral interferometry (SRSI) \cite{rhodes2014standards}. Moreover, it has been shown that satellite pulses with variable separations and/or relative phases with respect to the main pulses are invisible when measured by spectral-interferometry techniques, such as SPIDER \cite{rhodes2017unstable}. Because FROG provides feedback on the measurement by comparing the agreement between the retrieved and measured traces, it was able to see unstable satellite pulses in all cases. In studies performed so far, it did not retrieve the correct relative pulse heights, but additional information in the measured trace could provide the correct heights \cite{rhodes2017unstable}.

Of course, there are infinitely many types of possible pulse-train instabilities, and they can be difficult to detect and identify. Here we consider one such, possibly common, situation:  when the group delay dispersion of the pulses changes rapidly, without concomitant change of the amplitude structure. Such a situation can arise if angularly dispersive elements, e.g., grating compressors or stretchers are used for a laser with pronounced beam pointing instability. It can, in principle, also arise when pulse energy fluctuations occur in a self-phase-modulating medium. This type of fluctuation is virtually invisible in autocorrelations. As a result, it is not generally discussed or of concern. \1{One may now argue that RF techniques might be able to detect such artifacts, and this is certainly true for situations with rapid pulse-to-pulse fluctuations of the pulse phase structure. Nevertheless, if phase fluctuations are relatively slow compared to the repetition rate of the laser, RF methods are prone to completely miss out on this artifact while they would clearly corrupt optical pulse characterization with their much longer averaging times.}

In this work, we find, among other results, that FROG yields the approximate average pulse in the train and provides clear indications of this instability. SPIDER can also detect this instability by structure in its trace's background. We further show that some measurements from the literature \cite{gallmann1999characterization} show possible chirp instability, which can most likely be explained by a disadvantageous setup of this early SPIDER apparatus. On the other hand, if such issues can be resolved, our study suggests that SPIDER has the potential to detect such degradations of the pulse-to-pulse coherence.

\section{Fringe contrast in interferometric pulse characterization measurements}
\label{sec:motiv}
Interferometric visibility is a direct manifestation of coherence, \textit{i.e.}, the maximum visibility occurs when $\Gamma=1$. SPIDER relies on an interference pattern formed by two frequency-sheared pulses \cite{iaconis1998spectral}, and thus, to some extent, could be similarly interpreted. A SPIDER trace can be mathematically written approximately as
\begin{equation}
I_{\rm SPIDER}(\omega) = \left\langle \left| \tilde{E_j}(\omega) + \tilde{E_j}(\omega + \Omega){\rm exp}(i\omega T) \right|^2 \right\rangle_j,
\label{eq:spider}
\end{equation}
where $\Omega$ is the frequency shift between the two up-converted signals and $T$ is the delay between the two ancilla pulses. The SPIDER technique is further discussed in Section \ref{sec:chirpins}.  When using an ensemble of pulses with varying temporal structure, the main effect on a SPIDER trace is the degradation of fringe visibility \cite{ratner2012coherent,rhodes2013pulse}. In the presence of a simple coherent artifact, this loss of fringe visibility is homogeneous across the entire SPIDER trace.

\begin{figure}[tbh]
\centering
\includegraphics[width=0.8\linewidth]{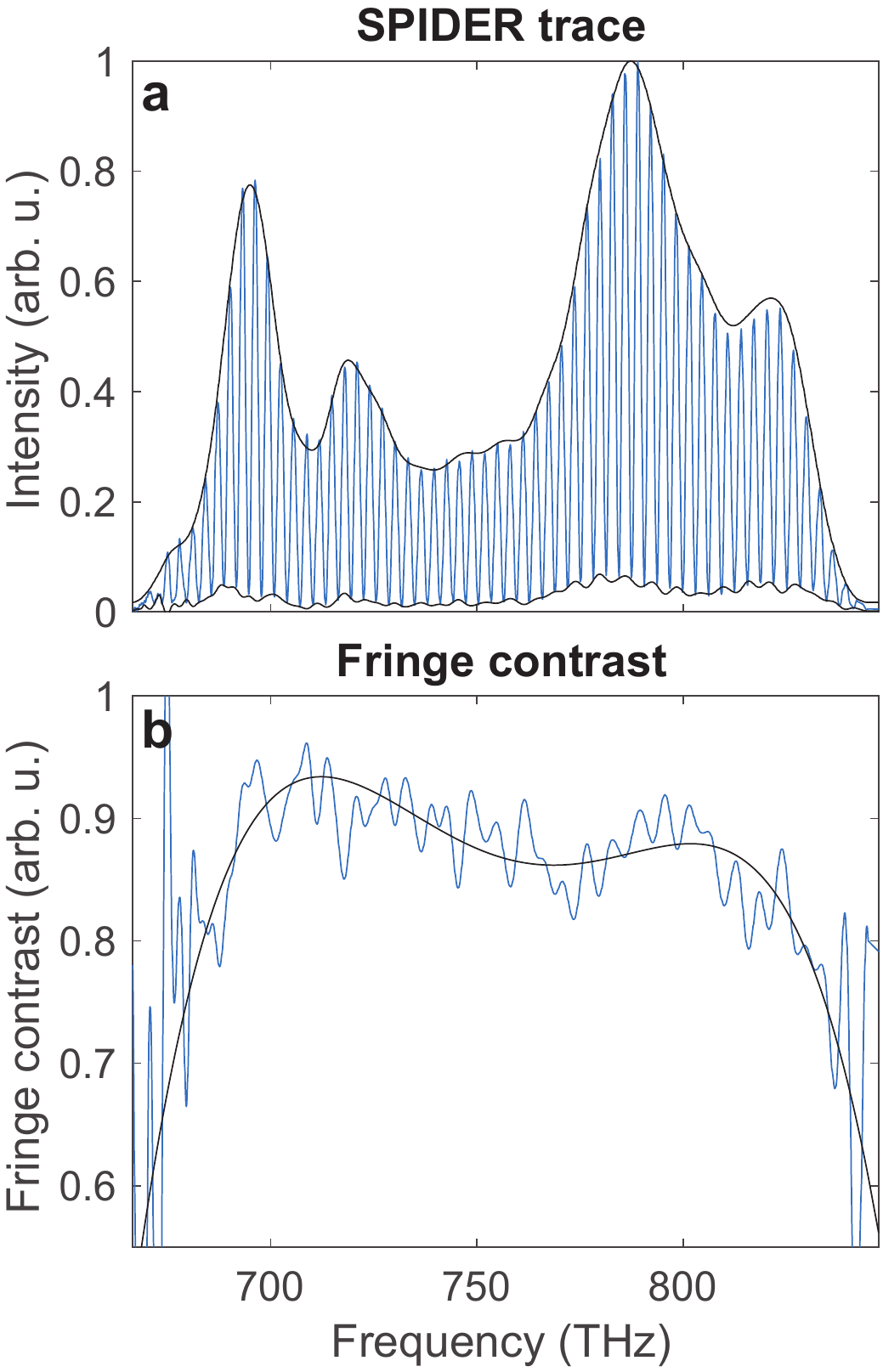}
\caption{(a) SPIDER trace from \cite{gallmann1999characterization}, with the maximum and minimum intensities traced by black lines. (b) Fringe contrast of the SPIDER trace in (a), with a polynomial fit in black to emphasize the general trend.}
\label{fig:ethdata}
\end{figure}

In terms of reduced fringe contrast, an interesting case is shown in Fig.~\ref{fig:ethdata}, which is the first measured SPIDER trace for sub-10-fs pulses \cite{gallmann1999characterization}. Here the visibility is high near the center but substantially decreases in the spectral wings of the trace. This rather localized reduction of fringe visibility differs from the expectations described in \cite{ratner2012coherent} for the presence of a coherent artifact, suggesting that a different mechanism is at work here. The most straightforward explanation is spectral reflectivity variations of the beam splitter coatings, but this can be rather safely ruled out here, as a reflection balanced scheme was employed in \cite{gallmann1999characterization}. As we further analyze below, one possible explanation of the parabolic contrast reduction in Fig.~\ref{fig:ethdata}(b) is a pulse-to-pulse variation of the linear chirp, and in fact, the laser under investigation in \cite{gallmann1999characterization} contained a prism sequence \cite{Fork1984}, which may have translated beam pointing fluctuations into chirp variations.

\section{Chirp instability}
\label{sec:chirpins}

For a systematic study of the effects of this instability on different techniques, five trains of 5000 pulses each were simulated. Each pulse has the same spectrum as that of the pulse that yielded the SPIDER trace in Figure~\ref{fig:ethdata}. The spectral phase of each pulse is allowed to vary, according to a normal distribution of linear chirp with a mean of zero. The standard deviation $\sigma$ of the \1{group delay dispersion (GDD) or chirp} for each train is different, uniformly distributed from zero to 60 $\rm fs^2$. The one with $\sigma = 0$ represents a stable train of Fourier-limited pulses.

\begin{figure}[tbh]
\centering
\includegraphics[width=0.8\linewidth]{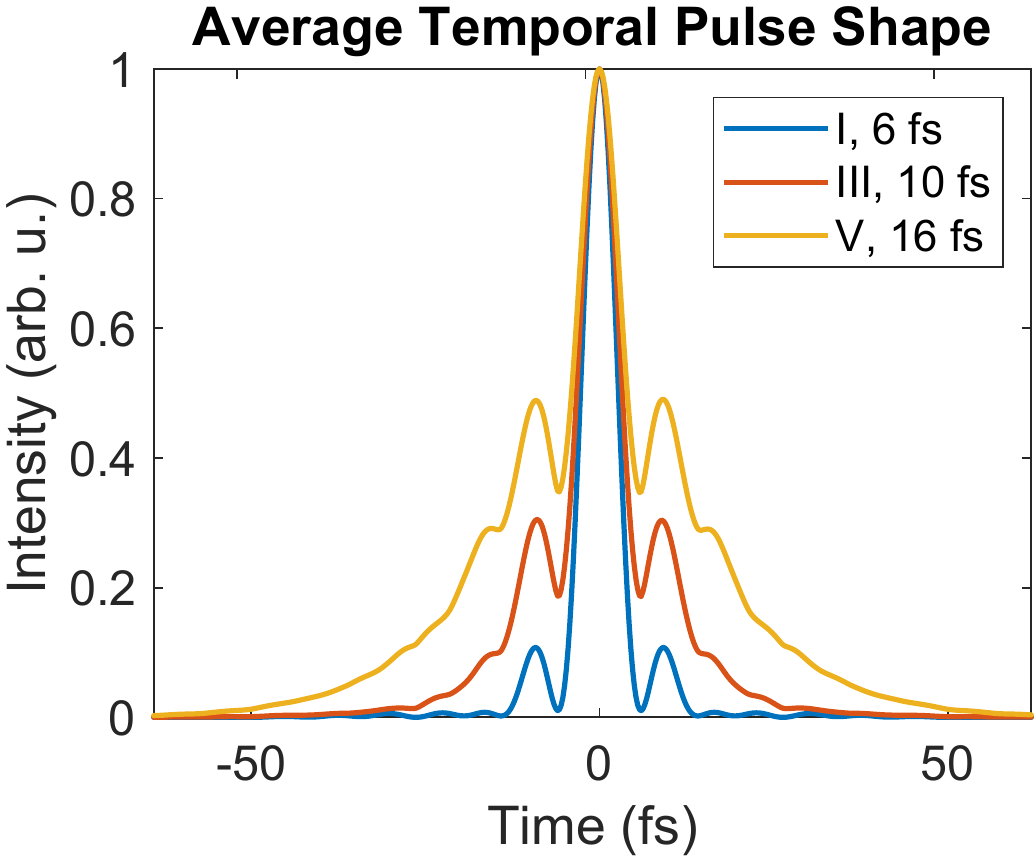}
\caption{\1{Average temporal pulse shapes for three of the ensembles. The temporal widths reported in the legend are the rms widths. The standard deviations of the GDD variation in the ensembles are: I, 0 $\rm fs^2$; III, 30 $\rm fs^2$; and V, 60 $\rm fs^2$.}}
\label{fig:temporal}
\end{figure}

\1{Figure~\ref{fig:temporal} displays the average pulse shapes of three of the ensembles, showing the influence of chirp instability. The rms width of the average pulse shape more than doubled in the presence of the largest amount of chirp instability we introduced. Yet, the full width at half maximum is barely affected, since most of the expansion is in the wings of the pulse.}

The first method we considered is intensity autocorrelation, which can be described by
\begin{equation}
I_{\rm ac}(\tau) = \left\langle \int^\infty_{-\infty} \left|  E_j(t)E_j(t - \tau)  \right|^2 {\rm d}t \right\rangle_j,
\label{eq:autoc}
\end{equation}
where $E(t)$ is the electric field in time domain, and $\tau$ is the delay between the two replica pulses.
\begin{figure}[tbh]
\centering
\includegraphics[width=0.8\linewidth]{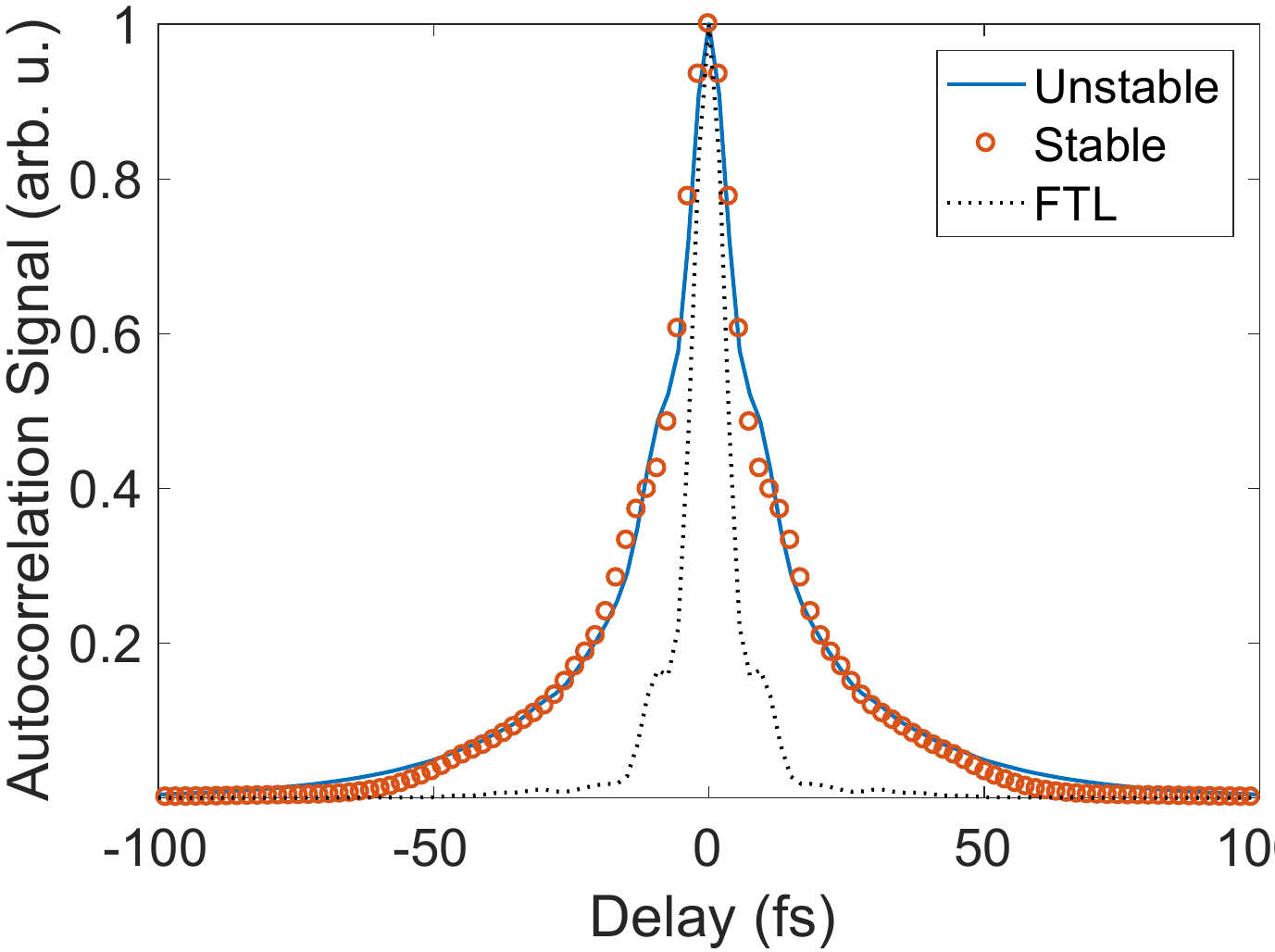}
\caption{Autocorrelation measurement for $\sigma = 60 \rm fs^2$ (blue line), along with the autocorrelation trace of a stable pulse, for the given spectrum accompanied by higher order spectral phases (red circles), having a similar autocorrelation, and autocorrelation of the Fourier transform limited pulse, $\sigma = 0 \rm fs^2$ (black dotted line).}
\label{fig:autocor}
\end{figure}

\begin{figure}[tbh]
\centering
\includegraphics[width=0.7\linewidth]{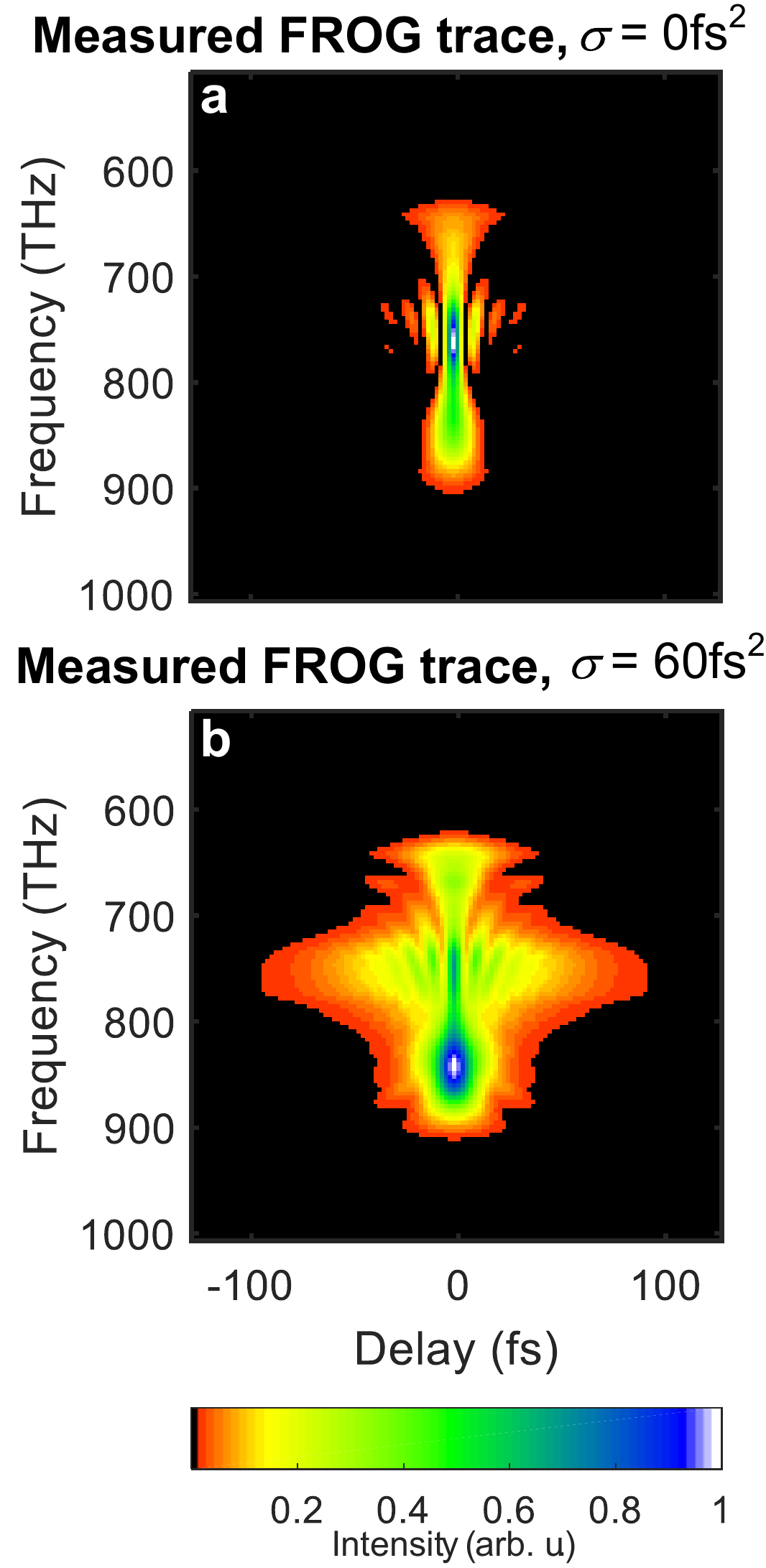}
\caption{FROG traces for the pulse trains with (a) $\sigma = 0$ and (b) $\sigma = 60 \rm fs^2$. Chirp variation causes a noticeable widening of the trace along the delay axis.}
\label{fig:measuredfrog}
\end{figure}

Because intensity autocorrelation is independent of the temporal phase, it only sees the variations in the intensity, here the pulse length.  The autocorrelation measurement for the pulse train with the largest chirp instability is shown in Fig.~\ref{fig:autocor}. The autocorrelation of a pulse with the same spectrum fitted to the signal is also plotted. Evidently, the autocorrelation of the pulse train with the largest GDD variation fits almost exactly to this stable pulse train, almost without any indication of instability. No coherent artifact appears, and this problem would remain completely undetected in autocorrelation measurements.  As a result, we do not consider autocorrelation further.

Spectrally resolving the autocorrelation gives us second-harmonic generation (SHG) FROG \cite{trebino2012frequency}
\begin{equation}
I_{\rm FROG}(\omega,\tau) = \left\langle \left|  \int^\infty_{-\infty}  E_j(t)E_j(t - \tau) {\rm exp}(-i\omega t)  {\rm d}t \right|^2 \right\rangle_j.
\end{equation}
Due to the uniqueness of a pulse retrieved from a FROG trace (except for some well-known trivial ambiguities) \cite{trebino2012frequency}, it is possible to retrieve the complex field from the FROG trace using iterative algorithms. FROG and its many variations are prevalent techniques for characterizing ultrashort pulses, but, in this work, we consider only the SHG version of FROG due to its popularity. Also, it is the weakest version of this class of powerful techniques in that it has an ambiguity in the direction of time, which other versions do not, so it would be expected to yield the poorest behavior of al the FROG variations in our study, and the other versions will likely perform better.

We find that, in an SHG FROG measurement, chirp instability causes a noticeable widening of the spectrogram along the delay axis, cf. Fig.~\ref{fig:measuredfrog}, similar to that seen in the autocorrelation measurement. But it also shows additional distortions, not visible in autocorrelation and unique to the case of chirp variation studied here.

\begin{figure}[tbh!]
\centering
\includegraphics[width=\linewidth]{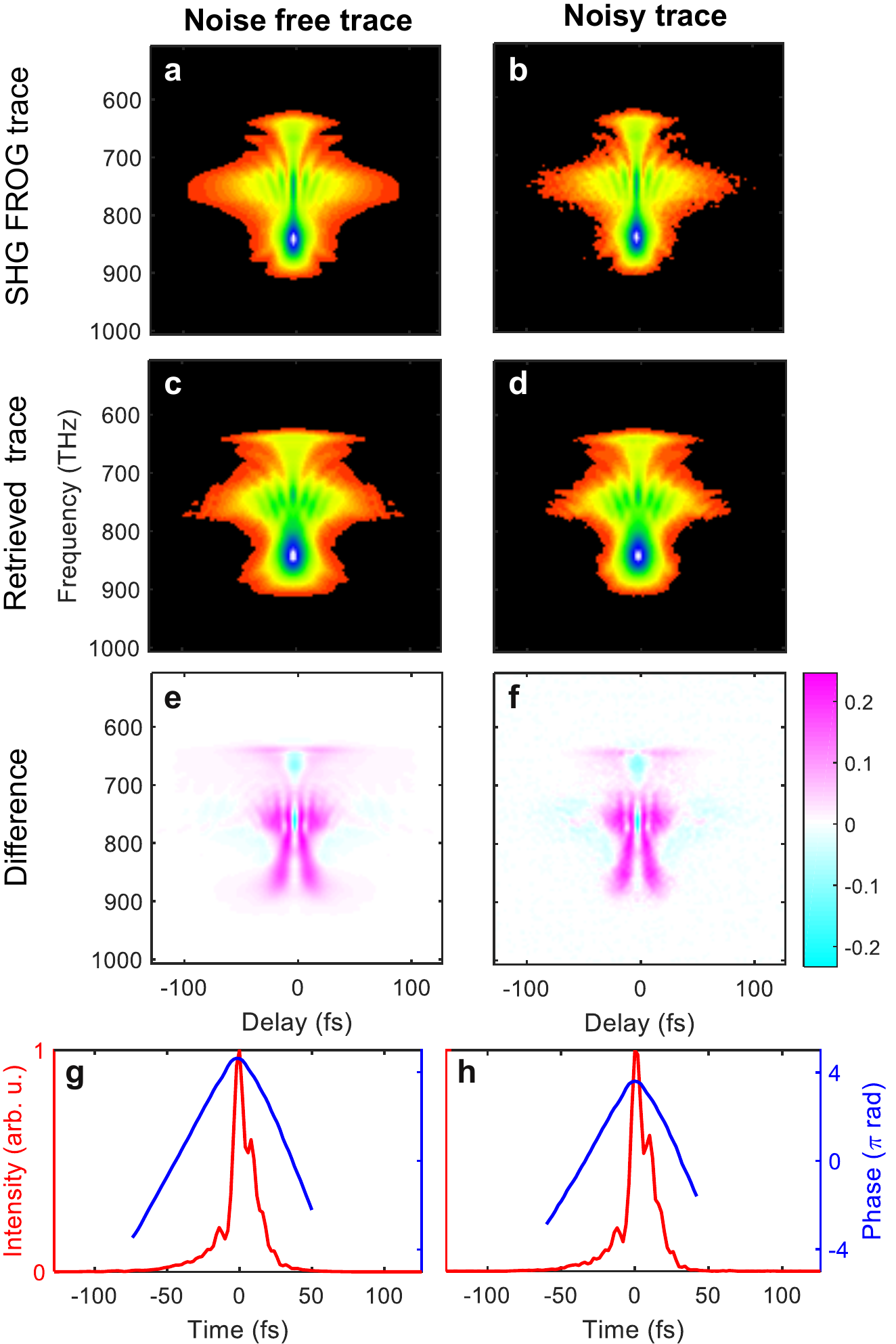}
\caption{(a, b) 128 $\times$ 128 noise free and noisy traces corresponding to $\sigma = 60\rm fs^2$, respectively. (c, d) Retrieved FROG traces ($G$ = 2.10\% and 2.13\%, respectively, which indicates poor agreement). (e, f) The difference of measured and reconstructed traces. A characteristic pattern is due to chirp instability. (g, h) Retrieved fields with rms widths 16.8fs (g) and 15.1fs (h), close to the average width of pulses in the train (16.1fs). }
\label{fig:retrievedfrog}
\end{figure}

As in other cases of unstable pulse shapes and FROG, attempting to retrieve the complex electric fields from the traces corresponding to the given unstable pulse trains results in retrieved traces that do not agree with the measured ones. This is readily revealed by the relatively large value of $G$ error, the rms difference between the measured and retrieved traces, for 128 $\times$ 128 noise-free SHG trace corresponding to $\sigma = 60\rm fs^2$ ($G$ = 2.10\%, cf. Fig.~\ref{fig:retrievedfrog}). Additionally, the difference map can (and should) be used to detect the presence of chirp instability. This discrepancy is revealed by the distinct pattern in the difference between the retrieved trace and the measured trace, as shown in Fig.~\ref{fig:retrievedfrog}(e, f). The difference between the traces is significant and non-random, indicating significant pulse-train instability, here in the magnitude of linear chirp. This discrepancy exceeds usual experimental noise levels and has a distinct symmetry along both delay and frequency axes, which allows it to be distinguished from experimental errors. Based on the retrieved electric fields, as shown in Fig.~\ref{fig:retrievedfrog}(g, h), SHG FROG yields the average rms widths of the train. Thus, SHG FROG yields the approximate average pulse in the train and provides clear indications of instability.

The dispersion scan (d-scan) method is very similar to FROG, but measures second-harmonic spectra of a pulse train as a function of dispersion rather than of delay. The former can be accomplished by varying the insertion of glass wedges in a pulse compressor setup \cite{miranda2012simultaneous}. Iterative algorithms are similarly used to retrieve the complex field from the d-scan trace \cite{escoto2018advanced}. The technique has recently been gaining popularity in measuring near-single-cycle pulses, since these pulses are very sensitive to dispersion \cite{timmers2017generating,silva2014simultaneous,tajalli2016few}.

\begin{figure}[tbh]
\centering
\includegraphics[width=0.7\linewidth]{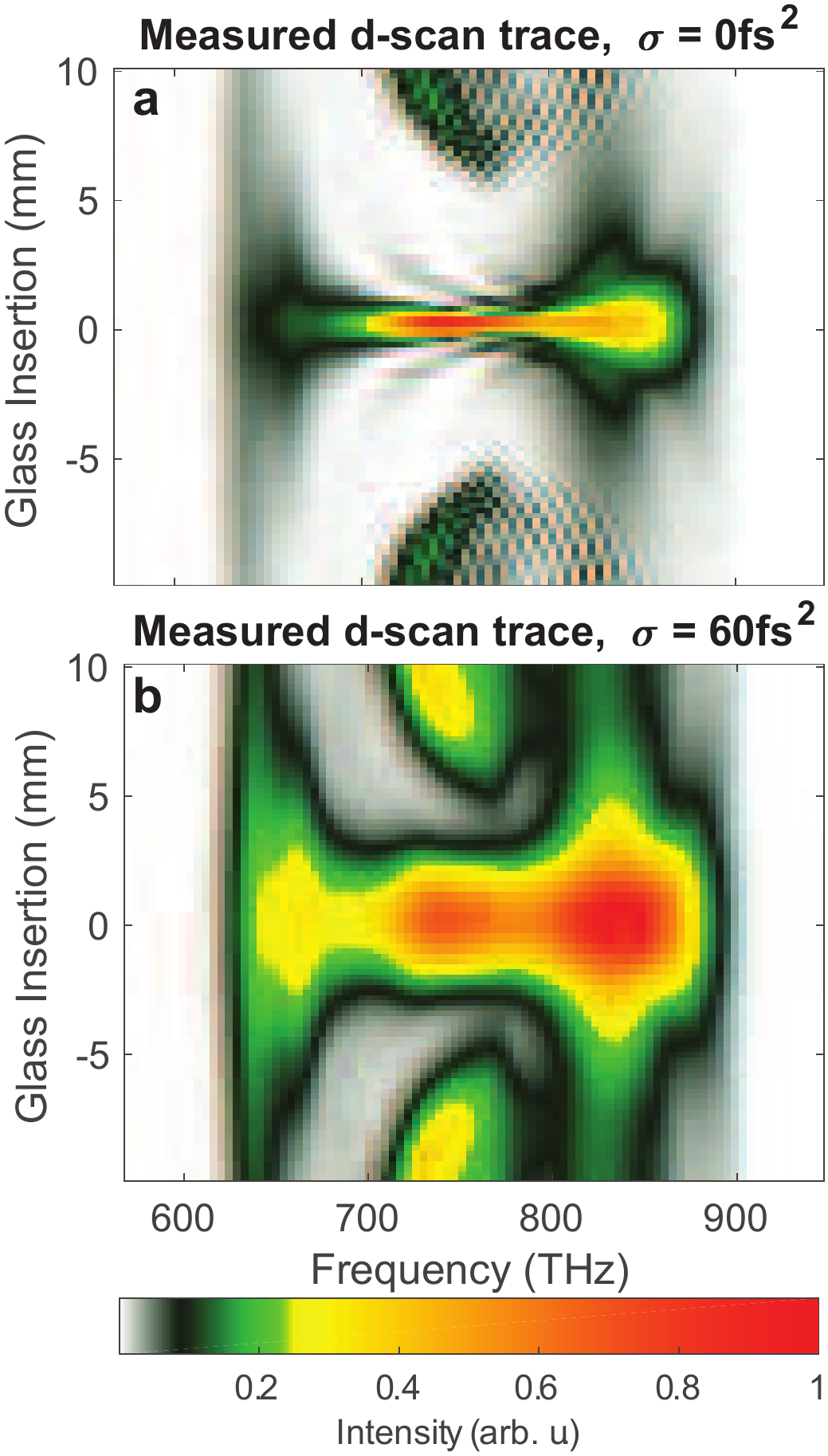}
\caption{D-scan traces for the pulse trains with (a) $\sigma = 0$ and (b) $\sigma = 60 \rm fs^2$. The trace spreads along the insertion axis due to chirp instability.}
\label{fig:dscan}
\end{figure}

A d-scan trace can be written as
\begin{align}
&I_{\rm d-scan}(\omega,z) = \nonumber \\
&  \Bigg\langle \Big|  \int^\infty_{-\infty} \left( \int^\infty_{-\infty} \tilde{E}(\omega) {\rm exp} (izk(\omega) + i\omega t) d\omega    \right)^2 \nonumber \\
&\times {\rm exp}(-i\omega t)  {\rm d}t \Big|^2 \Bigg\rangle,
\end{align}
where $z$ is the thickness and $k(\omega)$ is the frequency-dependent wave number of the dispersive material. Figure~\ref{fig:dscan} displays the traces for $\sigma$ of 0 and $60 \rm fs^2$. A d-scan trace in the presence of higher-order dispersion in the pulse usually shows shifting of the peak positions \cite{miranda2012simultaneous}, but not spreading of the signal along the insertion axis as can be seen in Fig.~\ref{fig:dscan}(b). We attempted to retrieve the phase from the latter trace, but the algorithm fails to find any meaningful single pulse shape that would give rise to the d-scan trace in Fig.~\ref{fig:dscan}(b). For a noiseless trace with $\sigma$ of $15\rm fs^2$, we already get a $G$ of 2.79\%, which goes up to 8.43\% for $60\rm fs^2$. We find that d-scan is quite sensitive even to small chirp instabilities and indicates their presence with large retrieval errors. \1{In particular, rather small $G$ errors on the order of a single percent may result in the retrieval, which are nevertheless an alarming indication of an underlying chirp instability. Similar conclusions can be made for FROG in the presence of a coherent artifact. While often ignored in literature, concomitant large retrieval errors $>1\%$ with simply structured FROG or d-scan traces appear to be valid indications of a coherence problem.}

We can go one step ahead and quantify the retrieval error by introducing a fidelity measure as originally suggested in \cite{lozovoy2015quantifying}. Fidelity is computed using the equation
\begin{equation}
F(z) = \frac{I_{\rm stable}(z)/I_{\rm stable}(0)}{\langle I(z) \rangle / \langle I(0) \rangle},
\end{equation}
where $I(z)$ is the measured intensity of the upconverted pulse at glass insertion $z$, and $I_{\rm stable}$ is computed using only a single pulse instead of an ensemble of varying pulses.

\begin{figure}[tbh]
\centering
\includegraphics[width=0.8\linewidth]{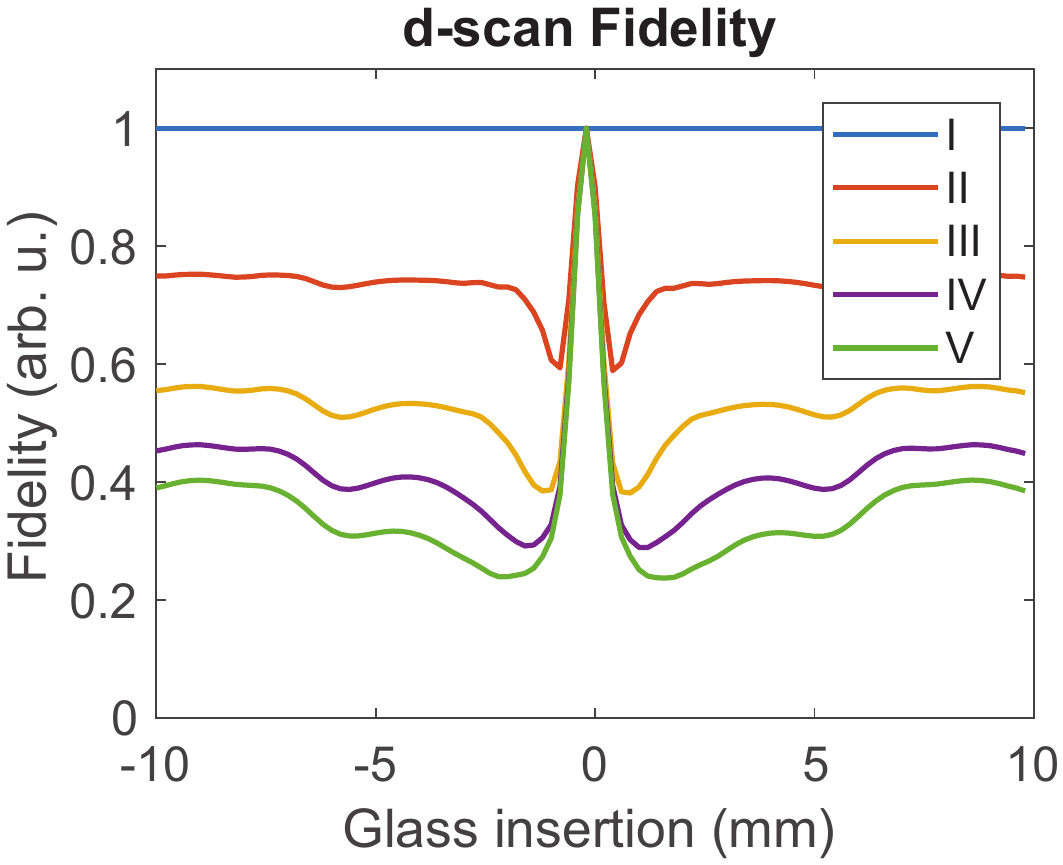}
\caption{Fidelity from the d-scan traces of pulse trains with different degrees of GDD variation, whose standard deviations are: I, 0 $\rm fs^2$; II, 15 $\rm fs^2$; III, 30 $\rm fs^2$; IV, 45 $\rm fs^2$; and V, 60 $\rm fs^2$. The dips on both sides near zero insertion are visible, which is a characteristic of spectral phase noise.}
\label{fig:fidelity}
\end{figure}

Figure~\ref{fig:fidelity} shows the fidelity plots for five different chirp variations. This analysis shows that a chirp instability yields d-scan traces that are structurally similar to those of unchirped pulses, yet are substantially elongated along the glass insertion axis. In the fidelity picture, this effect causes a characteristic drop of the fidelity with increasing distance from zero dispersion.

\begin{figure}[tbh]
\centering
\includegraphics[width=0.8\linewidth]{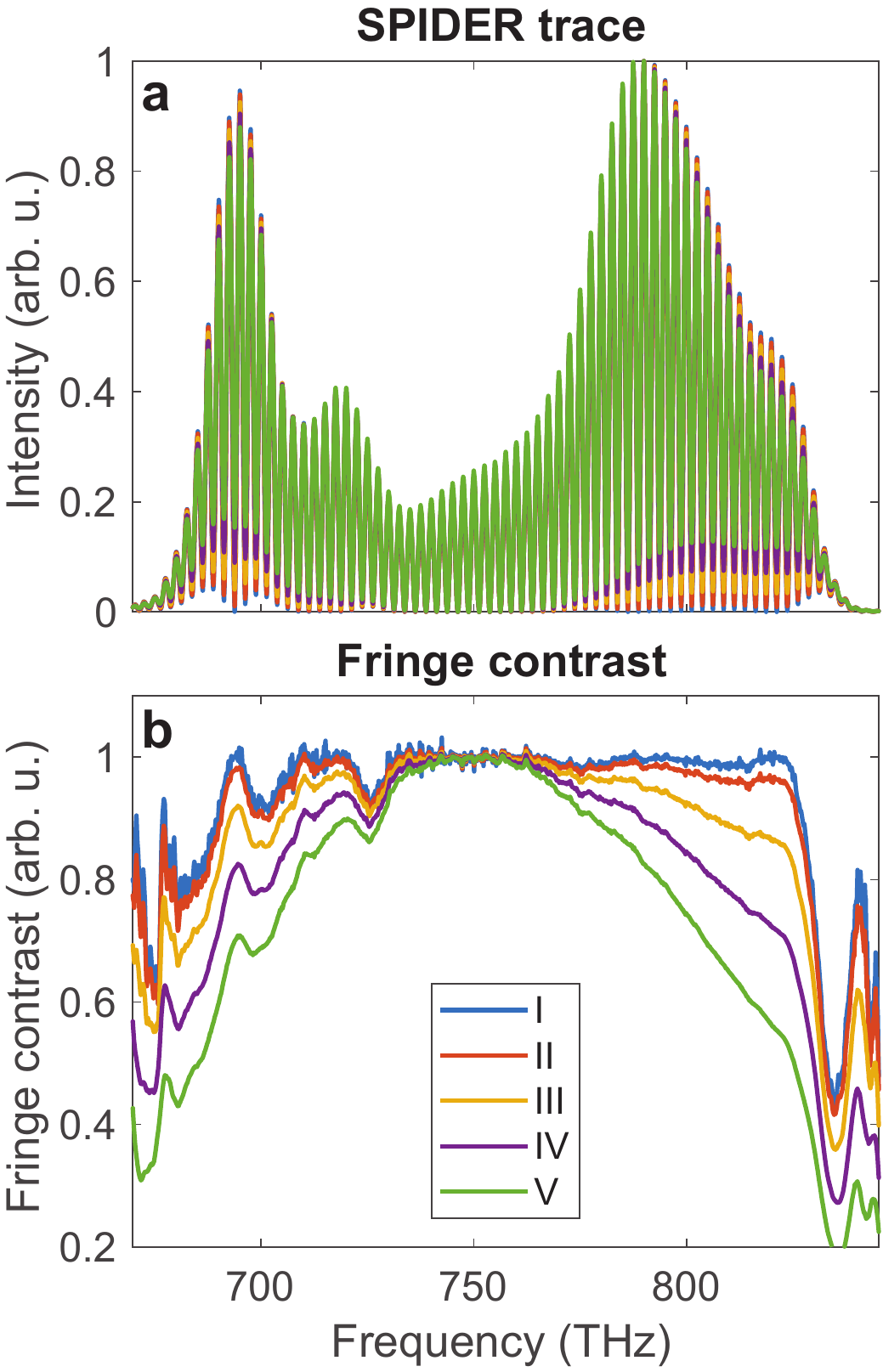}
\caption{SPIDER traces from pulse trains of different degrees of GDD variation, whose standard deviations are:  I, 0 $\rm fs^2$; II, 15 $\rm fs^2$; III, 30 $\rm fs^2$; IV, 45 $\rm fs^2$; and V, 60 $\rm fs^2$.}
\label{fig:spidersim}
\end{figure}

We next consider SPIDER. A SPIDER trace is an interference pattern formed by two frequency-sheared spectra of an ultrafast pulse. These sheared spectra are generated by sum-frequency generation between two delayed replicas of the pulse under test and a strongly chirped pulse. It is important to note that the SPIDER method requires the acquisition of a second interferogram for calibration purposes. This calibration measurement utilizes the second harmonic of the same replica pair. Figure~\ref{fig:spidersim} displays the SPIDER traces for the five different values of chirp instability. Even for the stable pulse train ($\sigma = 0$), the visibility is not perfectly flat, as it also depends on the intensity and phase of the pulse itself because of the frequency shear. On the other hand, the reference trace used for calibration exhibits a perfect visibility throughout.

Figure \ref{fig:spidersim}(b) shows a somewhat similar evolution of the fringe visibility in SPIDER as we previously noticed for the fidelity of d-scan traces, i.e., the fringe contrast is increasingly reduced in the wings of the spectrum, but is nearly perfect at the central wavelength. And this behavior matches the fringe contrast of the measured SPIDER trace in Fig.~\ref{fig:ethdata}. In the following, we discuss whether this fringe contrast reduction stems from the interaction between ancilla and replica pulses or is indicative of a chirp instability.

\section{Discussion}

In order to resolve the question of a potential chirp instability, the fringe contrast in Fig.~\ref{fig:ethdata} is compared with simulated SPIDER traces of the same pulse, assuming different amounts of chirp variation and an average spectral phase originally measured in \cite{gallmann1999characterization}. In our simulations, a Gaussian distribution of GDD with standard deviation $\sigma$ is assumed. The simulations clearly indicate that SPIDER measures the correct average spectral phase in the presence of chirp instability. This behavior is perfectly consistent with the findings in \cite{ratner2012coherent}, in which the spectral phase of each pulse in the train was a random complicated function, but whose average was a constant, yielding a very short retrieved pulse (much shorter than the actual average pulse).

We point out here that, in principle, overall, SPIDER should yield an accurate value for the average chirp when it is larger than its variation.  This is because, in general, SPIDER measures the average spectral phase, and the average over various quadratic curves (all with the same sign in this case) would be expected to be about the average of the quadratics.  However, when this condition is not met and the average chirp is near zero, both positive and negative chirps will be present, both lengthening the pulse and hence its average.  But the SPIDER measurement would yield the average spectral phase, which would then indicate near-zero chirp, and hence a shorter pulse than would in fact on average be present.

\begin{figure}[h!]
\centering
\includegraphics[width=0.8\linewidth]{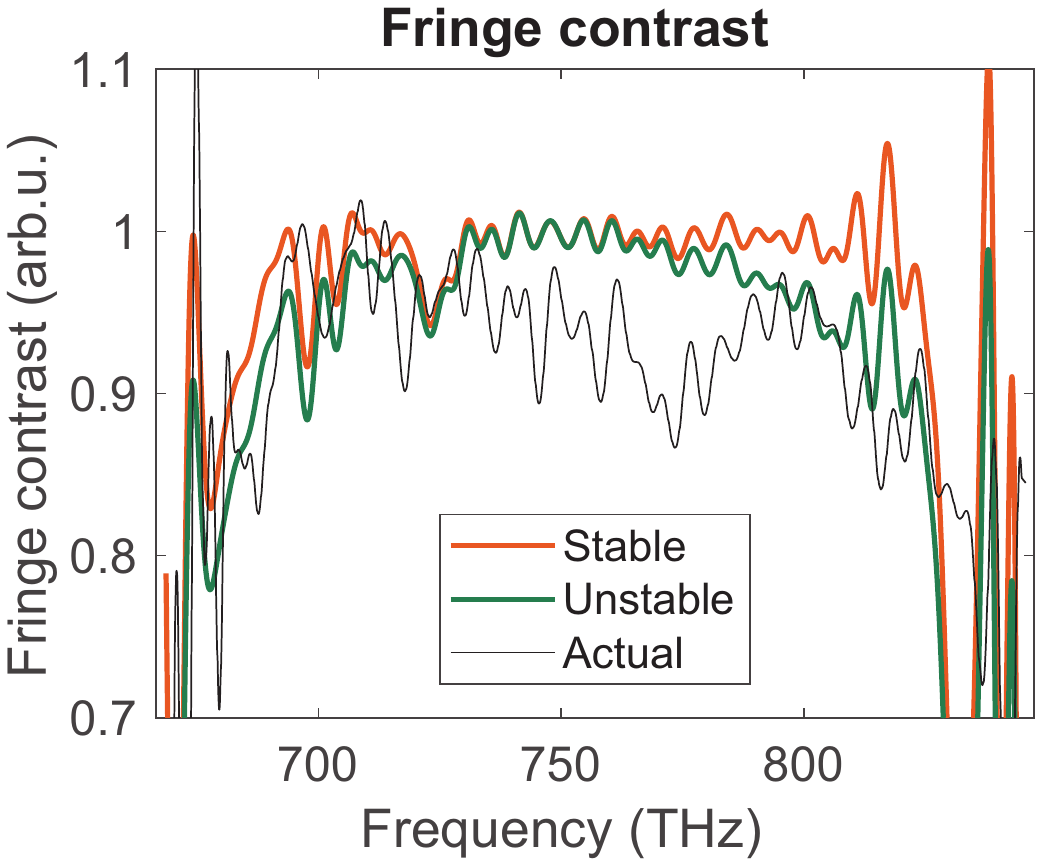}
\caption{Plots of the fringe contrast from three SPIDER measurements: stable ($\sigma = 0$), unstable ($\sigma = 29 \rm fs^2$), and actual (extracted from \cite{gallmann1999characterization})}
\label{fig:fringecontrast}
\end{figure}

\begin{figure}[h!]
\centering
\includegraphics[width=0.8\linewidth]{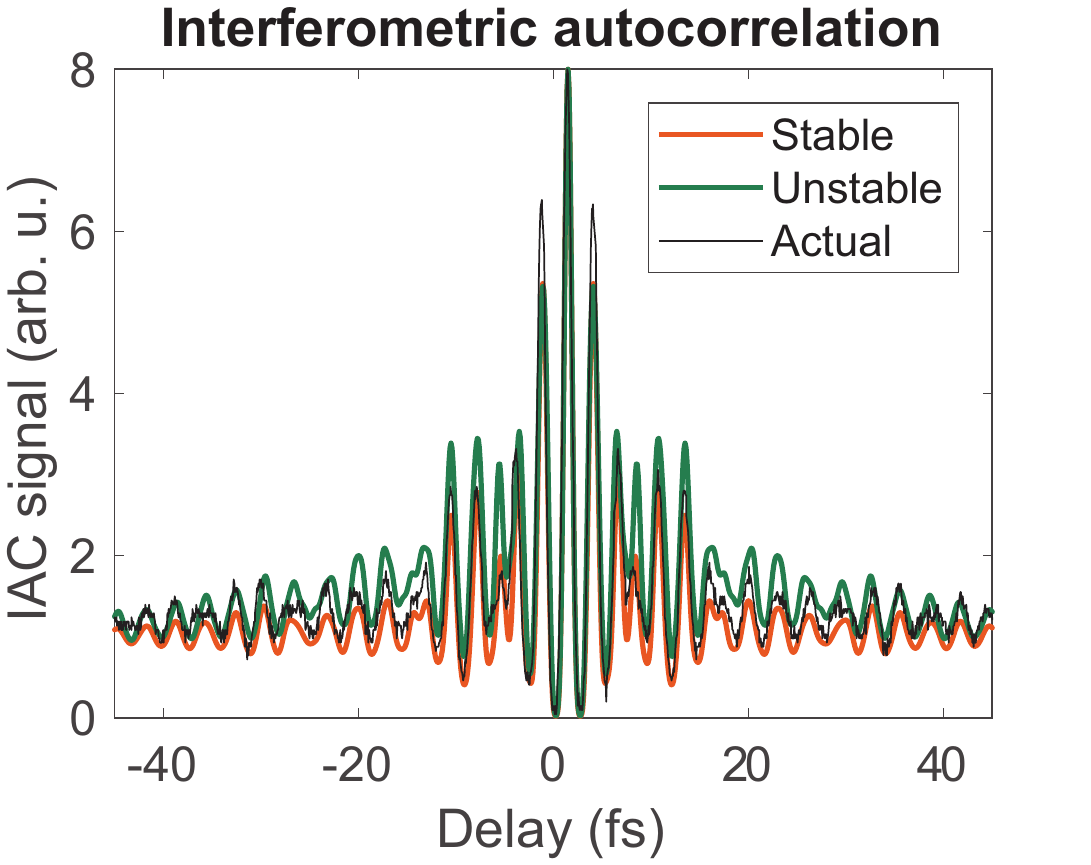}
\caption{Interferometric autocorrelation: stable ($\sigma = 0$), unstable ($\sigma = 29 \rm fs^2$), and actual (extracted from \cite{gallmann1999characterization}).}
\label{fig:iac}
\end{figure}

We found the best agreement between the originally measured SPIDER traces for a chirp variation of  $\sigma = 29 \rm fs^2$, see Fig.~\ref{fig:fringecontrast}. Taking into account that the simulated value of $\sigma$ should lead to a quite substantial broadening of the pulses by more than a femtosecond, one can independently verify the presence of a chirp instability by comparing to the autocorrelations that were also measured in \cite{gallmann1999characterization}, see Fig.~\ref{fig:iac}. This comparison clearly contradicts the presence of the chirp instability, as the pulse would have appeared markedly longer in autocorrelation measurements. Another way for such verification would have been the inspection of the fringe contrast of the calibration interferogram, but this measurement is unfortunately not available anymore.


We therefore conclude that the apparent chirp instability is actually caused by an artifact of the SPIDER setup used in Ref.~\cite{gallmann1999characterization}. There are several different ways of replica/ancilla preparation. In \cite{gallmann1999characterization}, a balanced Michelson interferometer was used. This balanced setup with two beam splitters compensates for small deviations from an ideal 50\% beam splitting ratio, which, in principle, allows to obtain near-ideal fringe contrast. However, at the same time, the Michelson setup is also prone to air turbulence. In order to evaluate a potential influence of resulting timing drift between the replica pulses, we repeated the simulations assuming a 1\,fs Gaussian timing jitter between the replicas and equally well reproduce the fringe contrast reduction of Ref.~\cite{gallmann1999characterization}. This timing jitter corresponds to a length variation of only about 150\,nm in one of the arms of the Michelson interferometer. This problem of few-cycle SPIDER setups has already been recognized more than a decade ago, \textit{i.e.}, ideally, the delay between the replicas must be maintained within few-attosecond precision \cite{Birge2006}. When the SPIDER technique advanced, one had consequently resorted to a different method for the replica preparation, using a single solid etalon to ensure sufficient delay stability \cite{stibenz2006optimizing}. However, this setup typically comes with lower fringe contrasts and is therefore less sensitive for a possible chirp instability.

\section{Conclusion and outlook}

We investigated the influence of a dynamic chirp instability on a number of well-established characterization techniques for few-cycle pulses. Compared to the widely explored coherent artifact, a chirp instability is typically more difficult to detect and easily overlooked in autocorrelation measurements.

SHG FROG measures a pulse that is close to the average pulse shape and width, and a simple post-retrieval comparison of measured and retrieved traces can check if this instability is present. FROG and its variations have been able to distinguish all forms of instability tested for so far.  Its two-dimensional trace allows for a broad range of indicators of different types of instability.  In each case, it provides a reasonable pulse length and has characteristic trace discrepancies that can be used to identify the type of instability (\textit{e.g.}, the usual coherence spike for partial mode-locking and washed out fringes in the central lobe for unstable multi-pulsing). And, with some additional effort, its magnitude can be usually be found.

Also, d-scan appears to be sensitive for this artifact, resulting in large retrieval errors even with the slightest instability we tested. Similar to FROG, it benefits from having a two-dimensional trace, which allows counter-checking of the reliability of the retrieval. Quantifying the instability is also possible using the same techniques presented in \cite{gerth2019regularized}.

Most surprisingly, SPIDER also seems to be remarkably sensitive to chirp instability, at least if delay variations between the replicas can be suitably excluded. As they showed the characteristic hallmark of this instability, we re-analyzed some nearly twenty-year-old measurements published in \cite{gallmann1999characterization} and found that the observed fringe contrast reduction could either be explained by a $29\rm fs^2$ group delay dispersion jitter on the pulse train or by delay variations in the replica preparation. Additionally, analyzing the autocorrelation data of this publication, we decided that the fringe contrast variation is most likely explained by the latter, \textit{i.e.}, atmospheric turbulence in the Michelson interferometer that was used at the time.

In summary, our investigations indicate new approaches for the safe detection of such very well-hidden instabilities. One such way is the careful analysis of the fringe contrast of the SPIDER measurement and the reference interferogram that is simply measured from the SHG of the two replicas. Optimizing the latter for high fringe contrast, even small chirp variations might become measurable from fringe contrast reductions in the SPIDER trace. Numerical simulations then allow to also reconstruct the average pulse shape in the time domain, \textit{i.e.}, one has access to resulting pulse length variations, too.

Finally, we see promise in the d-scan technique for unveiling mode-locking instabilities. As it relies on a one-beam geometry, it cannot be corrupted by time-delay variations as SPIDER obviously can. Moreover, this method has been successfully used for characterizing some of the shortest pulses generated to date. As our simulations clearly show, d-scan is not easily fooled by an underlying dynamic instability. Even rather small chirp fluctuations or the presence of a weak coherent artifact immediately results in distorted d-scan traces that result in large errors and erratic behavior of the retrieval algorithm. This further confirms the capabilities of d-scan to clarify situations of unstable mode-locking \cite{gerth2019regularized}.

In closing, we think that there is hope that we can finally overcome the problem of misinterpreting unstable mode-locking. Not only can we detect various artifacts based on the more thorough inspection of SPIDER, FROG, or d-scan traces, it also seems to be possible to simultaneously determine the average waveform together with its statistical spread, both in the time or frequency domain. Apart from laser oscillators with extremely short upper-state lifetime, this also serves some urgent needs in the endeavors of compressing pulses down to the single-cycle limit and possibly below.

\section*{Funding Information}
Deutsche Forschungsgemeinschaft (DFG) (STE 762/11-1); National Science Foundation (NSF) (\# ECCS-1609808); the Georgia Research Alliance.

\section*{Acknowledgments}
GS acknowledges fruitful discussions with Ian Walmsley (University of Oxford) about the fringe contrast in SPIDER measurements.

\bibliography{chirp_insta}

\end{document}